\documentclass[preprint,11pt]{elsarticle}

\usepackage{amssymb}
\usepackage{xcolor}

\usepackage{multirow}
\usepackage{array}

\journal{}

\begin{document}

\begin{frontmatter}



\title{Privacy Concerns Raised by Pervasive User Data Collection From Cyberspace and Their Countermeasures}


\author[label1,label2]{Yinhao Jiang\corref{cor1}}
\author[label1,label2]{Ba Dung Le}
\author[label1,label2,label4]{Tanveer Zia}
\author[label3]{Praveen Gauravaram}
\cortext[cor1]{Corresponding author}
\address[label1]{School of Computing and Mathematics, Charles Sturt University, NSW, Australia}
\address[label2]{Cyber Security Cooperative Research Centre, Joondalup, WA, Australia}
\address[label3]{Tata Consultancy Services, Australia}
\address[label4]{Center of Excellence in Cybercrime and Digital Forensics, Naif Arab University for Security Sciences, Saudi Arabia}

%\newpageafter{author}
\begin{abstract}
The virtual dimension called `Cyberspace' built on internet technologies has served people's daily lives for decades. Now it offers advanced services and connected experiences with the developing pervasive computing technologies that digitise, collect, and analyse users' activity data. This changes how user information gets collected and impacts user privacy at traditional cyberspace gateways, including the devices carried by users for daily use. This work investigates the impacts and surveys privacy concerns caused by this data collection, namely identity tracking from browsing activities, user input data disclosure, data accessibility in mobile devices, security of delicate data transmission, privacy in participating sensing, and identity privacy in opportunistic networks. Each of the surveyed privacy concerns is discussed in a well-defined scope according to the impacts mentioned above. Existing countermeasures are also surveyed and discussed, which identifies corresponding research gaps. To complete the perspectives, three complex open problems, namely trajectory privacy, privacy in smart metering, and involuntary privacy leakage with ambient intelligence, are briefly discussed for future research directions before a succinct conclusion to our survey at the end.
\end{abstract}


% \begin{highlights}
% \item We categorise the developing pervasive computing environment in the context of user privacy.
% \item We identify the impending privacy concerns and study them in featured scenarios.
% \item We survey the state-of-art countermeasure technologies.
% \item We discuss complex privacy problems and potential research questions.
% \end{highlights}


\end{frontmatter}


\section{Introduction}
Technologies, especially computing and communication technologies have significantly influenced our busy daily lives and created a virtual dimension to our lives.
	This virtual space, we call ``cyberspace", is usually described as a web of knotting communicating handsets and computers knotted together via a variety of networks.
	In cyberspace, instances can include an individual's avatars, smart devices, cloud computing portals, network gateways and so on.
	The interconnections between instances in cyberspace offer convenient platforms for information exchange, e.g., the internet.
	With the explosive growth of the internet from the end of the last century, we have experienced a fundamental transformation regarding how information can be created, acquired, disseminated, and used. 
	The internet continues its development by connecting embedded sensors, electronic tags on goods or freights, networked cameras, smart phones and vehicles, and almost all the things that are used on a frequent/daily basis.
	\color{black}{This latest development, including the Internet of Things (IoT) and wireless sensor networks, has raised a new concept of how people approach the cyberspace.}
	\color{black}{Considering more items are being connected to networks, most people now rely heavily on this computing environment and advanced data analytics.}
	%In other words, with more and more physical items connected to networks people now live their lives with ambient computing and network support that also may collect personal information at the same time.
	%interact with cyberspace by using the connected version of things, and receive benefits or network-enhanced experiences without any traditional network devices.
	%The collected data is then be analysed using machine learning technologies
	\color{black}{The resulting benefits encourage more connectivity to even more items and more connectivity opportunities from different types of networks towards a `hyper-connected world'} \cite{conti2017internet} \color{black}{where people can freely interact between the physical world and cyberspace with other people and resources.}
	%is trend of connecting more things by embedding network components and generating more connectivity opportunities by self-organised networks is presenting a hyper-connected world \cite{conti2017internet} that encourages people participating and enjoying communications anywhere and anytime with other people and machines as well as among machines.
	%each other andbetween each other and to other human beings or machines, and the benefits of communications between machines.
	
	\color{black}{However, these connected items also collect personal information and feed the sensitive data for further analysis.}
	\color{black}{New surveillance to private activities and unexpected personal information leakage have been generated, which creates security uncertainties and reshapes the digital personal privacy landscape.}
	%Unfortunately, this trend of hyper-connectivity \cite{wellman2001physical} creates new opportunities for data collection and re-shapes the landscape of personal privacy.
	The concept of privacy in cyberspace was established in 1997 \cite{westin1997whatever}, helping the cultivation and fast development of the Internet.
	Many regulations and bills have been proposed and issued for the protection of privacy.
	A common background that cyberspace \color{black}{serves users' calls as a distant and abstract oracle} was generally recognised.
	%can be disconnected easily with the willing of users.
	This may be the fact how cyberspace has been considered in the last two decades: we use a computer or handset to connect to cyberspace; we enjoy the provided services with full control; and cut the connection by simply closing the websites.
	Fortunately or not, due to the new connected items in users' proximity, cyberspace no longer remains passive to users' calls.
	%With the new data collections in a hyper-connected world, cyberspace remains passive to user's calls no longer.
	\color{black}{Thus, we need to examine how cyberspace activly collect users' information and how this information is used or exchanged in cyberspace. It becomes essential for us to learn what new privacy concerns and issues may happen in cyberspace and what the corresponding countermeasures are.}
	
	%//TODO
	%change
	%changes with 3 models
	%1. at traditional border
	%2. ubiquitous data collecting along with users
	%3. during the process of human data collecting
	%typical changes are focused on
	%followed by some future vision and complex cases
	\color{black}{Novel} technologies have changed how user information gets collected from users' manual input to more pervasively collection.
	The \textit{pervasive user data collection} is conducted by both connected software and hardware that monitors users' activities in cyberspace and the physical world.
	%embedded into devicesdeveloped to comprehensively and ubiquitously collect user information for data analytics and advanced services in cyberspace.
	%With these data collecting devices and equipments, 
	%In a hyper-connected world, organisations, companies and even individuals are provided a new way to collect data of users' activities in both cyberspace and physical world.
	For users' activities in cyberspace, advertisement cookies, Flash content and hyperlinked images quietly collect what users are browsing, what applications users are focusing on and the data indicating user preferences.
	Between cyberspace and the physical world, smart devices and wearable gadgets including mobile phones, smart watches and fitness trackers have been equipped with more resources so that they are capable of recording, storing and processing health data, daily routines, and other activity data. 
	Furthermore, for users' activities in the physical world, ubiquitous computing devices, especially the mobile phones, are now capable of collecting and uploading captured images, sounds, voices and videos, which may cause involuntary consent.
	The change of pervasive data collection triggers many new privacy concerns, and countermeasures are needed to prevent user privacy from slipping into a turmoil.
	
	The rest of this paper is organised as follows. In Section \ref{sec2}, we discuss privacy concerns at the traditional borders with cyberspace and survey two featured scenarios of user activities in cyberspace, namely web privacy protection and user input disclosure, including existing technologies. In Section \ref{sec3}, privacy problems between cyberspace and the physical world are investigated. This focuses on safe data accessibility and delicate transmission security. In Section \ref{sec4}, we explore user activities in the physical world involving pervasive computing where user privacy in participatory sensing and opportunistic networks riase user privacy concerns. Open issues are discussed in Section \ref{sec5} considering privacy protection for complex applications. The conclusions of this paper are outlined in Section \ref{sec6}.
	
	%In the following, we investigate privacy concerns and their current countermeasures in emerging scenarios around user activities from cyberspace to the physical world. 
	%We also discuss some complex privacy issues in future vision before the conclusion at the end.}
	%It can be seen that individual's privacy may slip into a turmoil if privacy protection does not cover sensors in the pervasive presence around us and mobile devices carried by people.

	\section{Privacy in data exhaust tracing}\label{sec2}
	First we \color{black}{take a close look at} the traditional border between the physical world and cyberspace, e.g., the gateways from web browsers. 
	
\begin{figure}
    \centerline{\includegraphics[width=28.5pc]{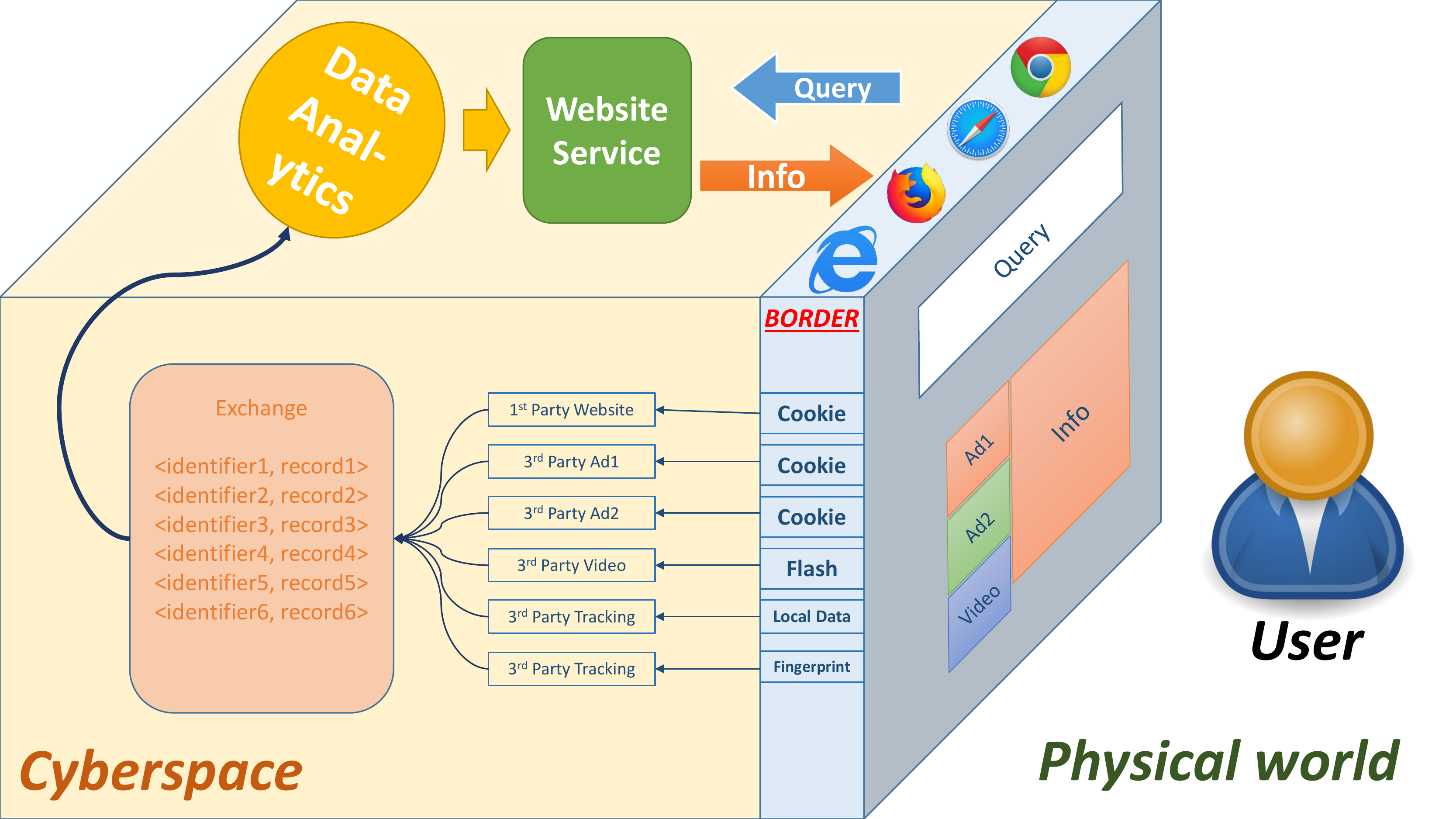}}
    \caption{Analysing and tracking activities at the border. User preference information gets collected, exchanged and analysed from browsers.}
    \label{fig:adcbsp}
    \end{figure}

	With the rise of machine learning applications, \color{black}{big data analysis has been developed on the data exhaust from people's daily we browsing.}
	\color{black}{The advertising industry, usually the third-party domains connected with publishers' websites, uses tools like cookies that put different identifiers on users so that browsing data exhaust can be traced and used to reconstruct individual browsing history.} Fig \ref{fig:adcbsp} illustrates that different tracking entities collect data exhaust underneath normal web browsing, using cookies \cite{zawadzinski_sweeney_2019}, flash \cite{soltani2010flash} or browser fingerprinting \cite{eckersley2010unique}, and analyse the data with cookie synchronising technology \cite{acar2014web}. 
    With the analysis, advertisers evaluate users' features including behavioural targeting, frequency capping, re-targeting and conversion tracking, to display their advertisements. 
    At the same time, the publishers also tailor information suiting users' conditions and predicted requirements based on evaluated preferences that the users have never set manually.
	
	The data exhaust collection and the following data analysis have brought incontrovertible benefits for both parties. While users enjoy the automatically personalised responses from a variety of websites and network services, publishers boast an approximately 52\% increased revenue from third-party cookie usage \cite{ravichandran_korula_2019}.
	However, this trend of data exhaust collection also triggers two major privacy concerns: identity tracking from browsing exhaust and private data disclosure. In the remaining part of this section, we survey these two privacy concerns and discuss solutions that can mitigate these privacy risks.

	\subsection{Identity tracking from browsing exhaust}
	At the traditional boundary between the physical world and cyberspace, advertising revenue has prompted the development of tracking individual browsing history, \color{black}{shopping behaviours and purchase habits} from browsing exhaust. 
	When a website is accessed, many entities other than the publisher domain are also connected. 
	The user generally classifies all the entities into two categories: first and third parties, where the publisher, the entity that the user visits is the first party,and other connected stakeholders, such as advertisers are the third parties that provide the publisher with a variety of services. 
	Generally, the adopted technologies from the third-parties are various cookies, local data storage \cite{soltani2010flash} or fingerprinting \cite{eckersley2010unique}, with which third parties put user identifications in temporary files and track from different publishers (or combine with HTTP referrer fields) to build a comprehensive data dump that will be later analysed for further evaluations or even exchanged with other third parties via cookie synchronising for an improved data collection.
	Thus, it becomes important that users' private information can be protected from ambient collection by third-parties.
	
	\subsubsection{Web privacy protection: browser extensions} Different techniques have been introduced to protect users from rowsing exhaust tracking.. Network monitor techniques include DNS filtering and network proxy, which are effective in specific scenarios but have apparent shortcomings against encrypted transmissions and individual URLs \cite{merzdovnik2017block}. 
	A popular type of protection comes from browser extensions that can reliably distinguish third-party content from a publisher's web pages and block unwanted content including encrypted web traffic. 
	In the following section, we describe several popular and featured browser extensions in detail. \color{black}{This information is summarised in Table} \ref{tbl:ext}.
	
	We classify selected tracking protection extensions by their blocking techniques: crowd-sourced lists, centralised maintenance, algorithm-based and machine learning based blocking.
	For extensions that are based on crowd-sourced lists, there are many famous and familiar names such as Ad-Blocker \cite{ad-blocker}, AdBlock \cite{adblock}, Adblock Plus \cite{adblock-plus} and uBlock \cite{hill_2020}. These extensions have drawn attentions in the community since the early days of the battle against third-party advertisements. As a result, a back-bone blocking list \textit{EasyList} has been contributed and maintained by the community. At the time of writing, \textit{EasyList} consists of over 17000 third-party advertisers, 13500 general, third-party and specific URL patterns as well as 31000 advertisement elements filters \cite{easylist}. With \textit{EasyList}, extensions like Ad-Blocker use filter rules to block ads from being loaded to help restrain user tracking. In addition to \textit{EasyList}, \color{black}{extensions based on rules also adopt other filter lists, including anti-circumvention lists or third-party tracker filter lists}. 
	The anti-circumvention lists help advertisement-blocking extensions fight against detection and circumvention to the extensions and re-insertion of ads. The third-party tracker filter lists help against tracking from companies and organisations that does not directly insert advertisements.
	
	The other popular type of advertisement tracking blocking relies on centralised maintenance including extensions like Ghostery \cite{ghostery}, Disconnect \cite{disconnect} and Blur \cite{abineblur}. Due to the centralised control, the extension companies set the blocking rules, and they typically have considerably fewer rules than crowd-sourced lists \cite{merzdovnik2017block}. 
	Although these extensions allow users to customise some blocking rules, certain network requests are defined as necessary third-party content, which cannot be blocked. Besides the difference in blocking ads, commercialised versions are provided with extra functionalities: Ghostery Insights \cite{ghosteryins} offers time-lined analysis and loading performance details; Disconnect Premium \cite{disconnectfq} offers an optional VPN, full IP masking and data encryption; Blur Premium \cite{abineincbp} offers further protections to personal information. \color{black}{Unfortunately, these new features requires not only financial support but also permission to collect user information}.
	
	An alternative type of protection extension, different to filter lists, uses algorithms to automatically decide whether a third-party's content needs to be blocked. A popular example is an extension called Privacy Badger \cite{eff_2018}, which counts how many websites a third-party organisation uses to track a user and blocks content from loading if the count for that organisation reaches three \cite{eff_2019}. In addition to blocking trackers from ads, Privacy Badger can also detect canvas-based browser fingerprinting and block tracking from the third-party domains \cite{eff_2019}.
	
    A newly emerged ad-blocking tool adopts machine learning techniques based on a perceptual study from the ads'loading content. \cite{storey2017future,tigas2019percival,paraska_2018} have introduced a new concept of perceptual ad-blocking, which seeks to improve resilience against ad obfuscation and minimise the manual effort needed to create ad-blockers. For traditional ad-blocking relying on crowd-sourced lists or ones based on centralised maintenance, two downsides have been identified: 1) the consistency of filter lists requires constant synchronisation with the latest versions; 2) different strategies have been developed for evading crowd-sourced lists (like \textit{EasyList}) such as changing domains, moving resources to the publishers, removing ad keywords from URLs, and removing image dimensions from URLs. Thus, it became an arms race between ads and tracker blocking tools and third-party domains. Researchers claim the novel approach of using perceptual signals effectively reduces the arms race with web publishers and ad networks \cite{tramer2019adversarial}. Storey et al. \cite{storey2017future} founded their perceptual ad-blocking based on a legal requirement for the recognisable display of ads by humans. Based on the legal requirement, Storey et al.'s Ad-Highlighter \cite{storey2017adh} focuses on learning captured visual and behavioural information that can be used to distinguish ads, e.g., the text ``Sponsored", the ad's circled ``i" information icon, or an ad network logo. However, their foundation is not as solid as they thought since the markup information, visual or behavioural, can be further rendered to an invisible state. To overcome the challenge of this specific rendering, \cite{paraska_2018} introduced the project Sentinel, a machine learning version of Adblocker Plus that uses an object-detection neural network to locate ads in raw website page screenshots \cite{adblockplusst}. To further exploit rendered web pages, \cite{tigas2019percival} introduced a new technique to achieve the goal. In their work, a deep-learning based ad-blocker module is embedded into Chromium's rendering engine so that images of ads can be detected directly  \cite{tigas2019percival}.
    
    Besides many experimental adoptions of machine learning for perceptual ad-blocking, \cite{iqbal2020adgraph} showed a different way of using machine learning based classification to block ads. Iqbal et al. \cite{iqbal2020adgraph} introduced AdGraph, which applies machine learning approaches on graph representations built from web pages considering aspects such as the HTML structure, network requests, and JavaScript behaviour. 
    A modification based on AdGraph was implemented on Chromium, resulting in high accuracy and fewer computational overheads compared to traditional ad-blocking extensions.

\begin{table}
\centering
\caption{Surveyed Web Browser Extensions}
\label{tbl:ext}
\small
\begin{tabular*}{27.5pc}{@{}|p{20pt}|p{93pt}<{\raggedright}|p{179pt}<{\raggedright}|@{}}
\hline
Ref.& 
Technique& 
Feature \\
\hline
\cite{adblock}&\multirow{7}{*}{Rule-based filter}&\multirow{4}{*}{Crowd-sourced list}\\
\cite{ad-blocker}&&\\
\cite{adblock-plus}&&\\
\cite{hill_2020}&&\\
\cline{1-1}
\cline{3-3}
\cite{ghostery}&&\multirow{3}{*}{Centralised maintenance}\\
\cite{disconnect}&&\\
\cite{abineblur}&&\\
\hline
\cite{eff_2018}&\multirow{2}{93pt}{Algorithm-based}&\multirow{2}{179pt}{Detect browser fingerprinting}\\
\cite{eff_2019}&&\\
\hline
\cite{storey2017future}&\multirow{4}{*}{Machine learning}&On image pattern\\
\cite{paraska_2018}&&On screenshot\\
\cite{tigas2019percival}&&On rendering engine\\
\cite{iqbal2020adgraph}&&On behaviour pattern\\
\hline

\end{tabular*}
%\label{tab1}
\end{table}

	\subsection{User input data disclosure}
	\color{black}{Another important change that happens at the border, which differs from the passive collecting and tracking activities, is how users' input gets recorded and utlised.}
	%usersprivacy issue is how data collected during web browsing and application activities will be used. 
	This change is associated with many questions simultaneously: how much personally identifiable input has been collected, how is the input data with sensitive information transmitted, how much sensitive information will be disclosed, would any critical information get leaked and would the user be re-identified from authorised analysis services, etc. 
	To cope with these questions, a relatively traditional approach, called differential privacy, was proposed that embeds extra noise and increases the entropy so that collected input data can be possibly denied by users and thus protected \cite{dwork2006calibrating}. 
	However, many problems remain for applying differential privacy technology to large-scale sensitive input data stored centrally \cite{yang2012differential}. 
	Conventional differential privacy is applied after data are collected in a centralised way and conducted using a quantitative approach by a trusted party. This brings uncertainties for users inclined towards independent intuitive methods when private information is involved. 
	To address the uncertainties, the recent topic of local differential privacy (LDP), where users randomly perturb their inputs to provide plausible deniability of their data without the need for a trusted party, has come to the fore.
	
	\subsubsection{Local differential privacy} The goal of differential privacy is to process a dataset in such a way that it is not possible to determine if a certain entry has been removed. The distribution remains indistinguishable in a preset range even if a certain entry got removed \cite{dwork2006calibrating}.
	When an algorithm achieves the goal with the preset range, it instantly guarantees that no observer can determine if a particular individual participated or if the data from this subject has been used\cite{35dwork2011differential}. 
	For LDP, users interact with an untrusted aggregator such that the aggregator learns statistical information about the distribution of the private value in the user population, while the information leakage for each individual is bounded.
	An important property for local differential privacy is that noise perturbation is conducted at the user end, so the collected data is not original.
	This property has brought significant applications in user data collection including {\it Google's RAPPOR} \cite{erlingsson2014rappor} and {\it Apple's Learning with privacy} \cite{apple_2017}.
	
	The foundation of most LDP realisations is an easily understandable concept extracted from {\it Random Response} \cite{warner1965randomized} that with a probability the collected data is the true value. 
	For different collecting styles and data use purposes, developments from {\it Google}, {\it Apple} and {\it Microsoft} are combined with other techniques targeting longitudinal collections, itemsets mining and repeated numeric value collection.
	
	{\it Google's RAPPOR} \cite{erlingsson2014rappor} uses a unary encoding that reports on a permanent randomised response for a question asked. 
	The technique ensures privacy, keeps information utility and helps with repeated data collection. 
	In the case of a survey with a large domain, they further adopted Bloom filters \cite{bloom1970space} for efficient element encoding.
	The {\it RAPPOR} technique has initialised a series of research developments in heavy hitter identification with LDP \cite{hsu2012distributed,bassily2017practical,wang2019locally}, which can be seen as an extended problem of frequency estimation \cite{wang2017locally}.
	
	After {\it Google's} efforts with heavy hitters, {\it Apple's} design focuses on the problem of frequent itemsets mining on a large scale.
	They first used discrete Fourier Transformation to cope with privacy sensitivity and noise adding, which successfully handles the sparsity of the input vectors.
	They then adopted sketching algorithms that reduce the dimensionality of the domain helping to learn the processed data.
	Apple's implementation successfully demonstrated an application of centralised differential privacy in the LDP setting.
	Several following works started a research trend on itemset mining \cite{wang2018locally} and marginal release \cite{cormode2018marginal} focusing on utilising LDP data.
	%Learing with Privacy: use sketching to reduce the dimensionality of a mssive domain itemset distinguish; using FT
	
	Although LDP has attracted much research from academia and industry, current research focuses on how to get useful information from collected LDP data. 
	An equally important question in LDP is how to implement LDP for other data collecting scenarios considering survey domain and approaches with specific characteristics.
	An example is {\it Microsoft's Telemetry collection} \cite{ding2017collecting}, which challenges a private numerical value scenario with very small but frequent changes.
	Other identified special survey domain includes graph data analysis \cite{qin2017generating}, language data analysis \cite{mcmahan2017learning}; and survey approaches considering multiple rounds of interactions \cite{nguyen2016collecting} and prior knowledge \cite{jia2019calibrate}.
	 
	\subsection{Discussion}
	We have surveyed protections on browsing exhaust and user response disclosure for data collection at the traditional border between the physical world and cyberspace. For browsing exhaust protection, there is an ongoing arms race between web browser extensions and exhaust tracking techniques. Core concepts for the developing browser extensions have evolved from elementary rule-based filters to perceptual blocking involving machine learning technologies, where further research can be focused. For private user response disclosure, LDP has shown its promising applications from {\it Google and Apple's}
	implementation. Despite the current research streams on LDP data utilisation, the realisation of LDP on special survey domains and approaches requires more research.

	\section{Privacy in personal data hub}\label{sec3}
    After close inspection of the changes happening at the border, we now investigate privacy concerns raised with the expanding industry of smart devices that collect private user data, store the data locally for access from applications, and transmit the data to the internet for further analysis and advanced services.
	
\begin{figure}
    \centerline{\includegraphics[width=28.5pc]{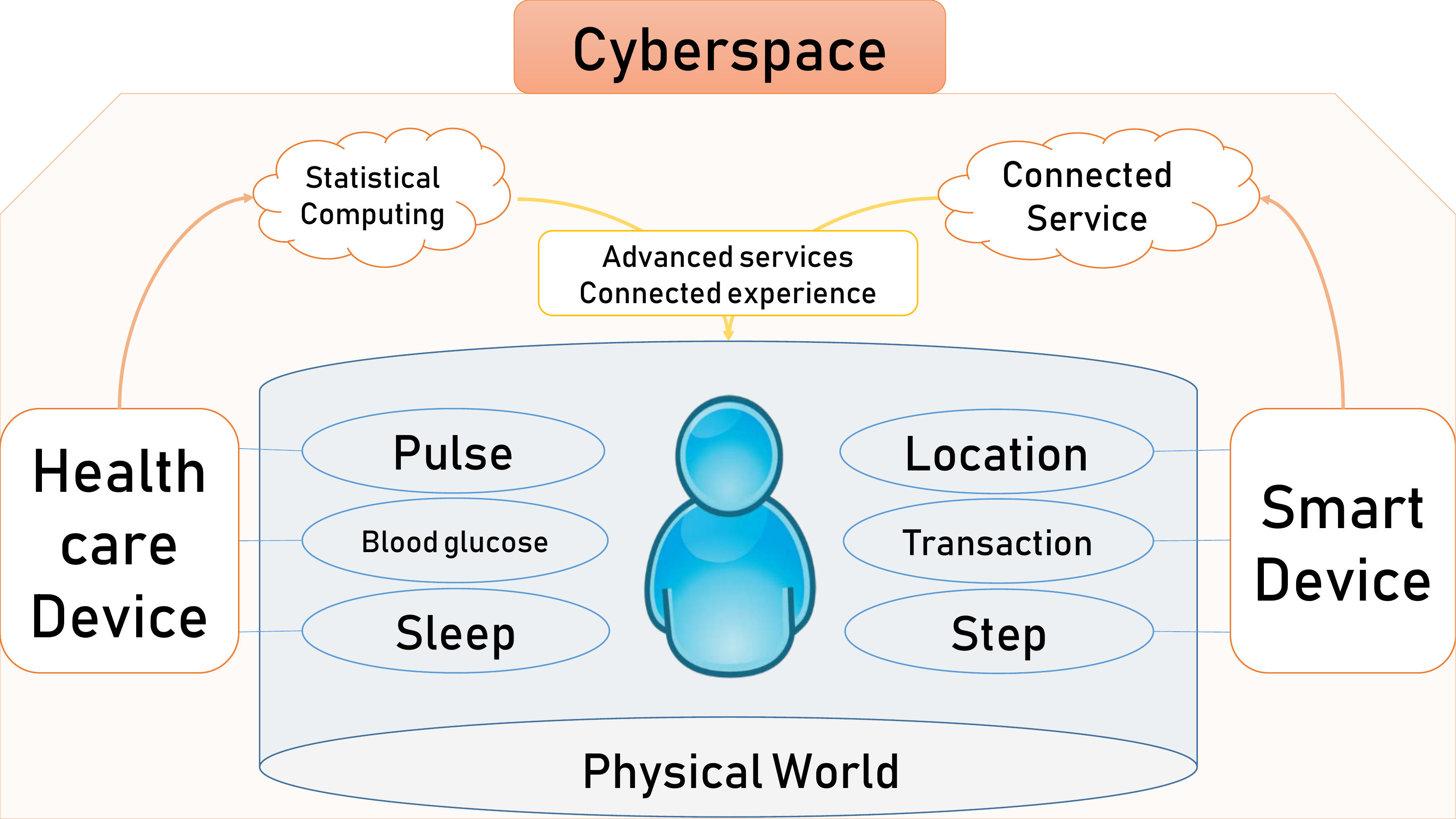}}
    \caption{Personal data hub paradigm. Smart devices and healthcare devices collect personal data for cyberspace entities to provide advanced services and connected experiences.}
    \label{fig:udr}
    \end{figure}

 	Smart devices and wearable gadgets can provide instant connected experiences and services with novel functionality.
 	At the same time, they also collect data and communicate with each other and the internet for further and better user benefits.
 	As showed in Fig. \ref{fig:udr}, the data collected by existing smart devices and wearable equipment concentrate on user status information that is considered highly private.
 	The collected user status information includes activity, location and health information, which can be used by different applications in smart devices and analytics from cloud computing.
 	Thus, the smart devices and wearable equipment form a personal data hub that collects, shares and distributes data from users, which leads to questions about user privacy protections.
 	
 	In such a personal data hub system, there are two important questions about user privacy: how to control the access of data stored locally and how to guarantee data security during transmission.
 	Although these two questions also exist in centralised data hub systems, the collected data in personal data hub systems mostly consists of personal information. Also, the devices carried by users do not have consistent computational resources due to heterogeneous manufacturer designs.
 	In the following section, we examine these two questions and state-of-art solutions.

	\subsection{Safe data accessibility}
	With smart devices and wearable gadgets playing important roles in people's health, an essential privacy problem is the collected data.
	These devices have a wide range of specifications regarding computing processors, memory and communication power.
	%For example, smartphones and high-end smartwatches often are equipped with high computational power and gigabyte level of memory.
	This brings them function space for third-party applications with direct connections to the internet.
	Thus, multiple entities could try to access these devices for their collected data, including malicious adversaries.
	In addition to potential external attacks, the collected data may contain highly sensitive information like health records and location coordinates, which could lead to misuse between applications on wearable gadgets and damage user privacy.
	
	\subsubsection{Access control enforcement for the wearable equipment}
	The technology of access control enforcement plays a core protection role in many IoT network systems, since it directly answers the privacy issue of accessibility mentioned above. 
	It applies a range of selective policies, setting the criteria of who can access the data. 
	The main purpose of an access control enforcement mechanism is to block unauthorised and random queries towards a protected data repository. 
	Besides the passive protection, rather than blocking arbitrary connections, it sets up a bottom line against insider attacks or general platform sharing with an efficient privilege update and revocation mechanism. 
	Access control mechanisms for IoT systems have drawn much research attention and several works have been proposed as effective and practical solutions for wearable technology in different scenarios.
	Since access control enforcement has a wide research scope, in the following section we survey a few typical works and focus on wearable gadgets and connected healthcare devices.
	
	One research focus required for wearable gadgets is to develop context-aware access control with a more expressive policy. In 2010, Garcia-Morchon and Wehrle \cite{garcia2010modular} proposed a modular context-aware access control mechanism that allows a system
    administrator to compose each module with a well-defined goal so that access policies for different required functionalities can be assigned to different modules. Ray et al. \cite{ray2017using} tried to improve the expressiveness using attribute-based access control from the NIST NGAC framework and achieved the first conceptual prototype for an IoT infrastructure. Later in the same year, Salama et al. \cite{118DBLP:conf/icws/SalamaYWPB17} successfully combined public key infrastructure and attribute-based access control for a multi-level access control on patient healthcare monitoring.
	%better policy
	
	Another research focus for wearable gadgets and connected healthcare devices is usability.
	This feature is neglected by most of the existing access control works since an administrative model is generally assumed for access control scenarios.
	However, especially for wearable gadgets, there is no administrative staff for these private devices and the users are the ones who configure, manage and protect the devices and resources.
	Thus, for the users who often lack the necessary security knowledge, an easy-to-use interface and enhanced presentation need to be provided for policy configuration \cite{ravidas2019access}.
	In 2011, Kim et al. \cite{kim2011access} proposed the first access control mechanism that provides a full solution to usability.
	Their newly introduced automated Clairvoyant access right assignment mechanism can suggest suitable access control policies.
	Unfortunately, their work is designed for smart home scenarios where its inherent \textit{overprivilege} property can be tolerated \cite{ravidas2019access}.
	To address this issue of overprivilege, Tian et al. \cite{tian2017smartauth} proposed an automated access policy generation based on checking the functionality and behaviour of the entity that asks for the access.
	Their access control mechanism was oriented for smartphone applications accessing local resources, which can be extended to other IoT systems like accessing data in wearable gadgets.
	After an appropriate access policy is generated, it is then provided to the user for review.
	
	Other research focuses include distributed environments \cite{alshehri2013secure,burnap2012protecting}, dynamic access control \cite{heydari2019towards}, scalability \cite{42rahman2017privacy} and multilateral security \cite{diez2019lightweight}. These works will be compared with aforementioned works from other focuses in Table \ref{tbl:acw}.
	
	\begin{table}
	\centering
    \caption{Comparison among access control enforcement for wearable equipment}
    \label{tbl:acw}
    \small
    \begin{tabular*}{27.5pc}{@{}|p{20pt}|p{285pt}<{\raggedright}|@{}}
    %\begin{tabular*}{17.5pc}{@{}|p{20pt}|p{165pt}<{\raggedright}|@{}}
    \hline
    Ref.& 
    Feature \\
    \hline
    \cite{garcia2010modular}&Modular context-aware access control\\
    \cite{ray2017using}&NIST NGAC framework\\
    \cite{118DBLP:conf/icws/SalamaYWPB17}&Multi-level access control\\
    \cite{tian2017smartauth}&Automated access policy generation\\
    \cite{alshehri2013secure}&BiLayer access control model\\
    \cite{burnap2012protecting}&Virtual patient record alternative\\
    \cite{heydari2019towards}&Indeterminacy-tolerant access control\\
    \cite{42rahman2017privacy}&Healthcare RFID tag access control\\
    \cite{diez2019lightweight}&Multi-level and multilateral security\\
\hline

\end{tabular*}
\end{table}
	
	\subsection{Security in delicate transmission}
	Besides guarding the access of the data stored locally, the problem of protecting sensitive data during transmission has brought many challenges to connected devices, especially resource-restrained healthcare devices.
	Solutions to this problem have been focused on the field of encryption and its efficiency optimisation. 
	While conventional encryption schemes like the Advanced Encryption Standard work fine on wearable gadgets with reasonable processing power and memory capabilities \cite{Centeno2018performance}, they meet various bottlenecks when applied to monitoring devices and small sensors. 
	To make the encryption approach compatible with these devices, lightweight encryption methods are proposed focusing on reducing computational overheads and increasing scheme efficiency.

	\subsubsection{Lightweight encryption in healthcare devices}
	Lightweight symmetric encryption can provide encryption requirements from connected healthcare devices, especially implantable medical devices like pacemakers where other protections are difficult to implement. 
	Connected healthcare devices are usually computationally weak and restrained by battery life, and implantable medical devices often are additionally restricted with a minimal physical size that leads to implementation constraints in hardware \cite{biryukov2017state,masoud2015power,ronen2017iot}.
	
	With these limitations, some features/properties in lightweight encryption become rather more acceptable and welcome.
	These features include implementation flexibility, smaller block size, encryption rounds saving, and restricted versatility.
	\begin{itemize}
	    \item {\it Implementation Flexibility--}For the implementation of encryption on resource-restrained devices, the trade-off is only determined when applied to a specific scenario \cite{rob16}.
	    Thus, when a feature is specifically needed for a deployment scenario, the encryption algorithm should be optimised with acceptable sacrifice to other aspects.
	    \item {\it Lower Size--}For healthcare devices that have a small physical size and need to run for an extended period with limited battery, the design of an encryption algorithm may need to prioritise resource limitations. In this case, a smaller size of block size or internal state becomes acceptable.
	    \item {\it Less Rounds--}For healthcare devices, a particular nature is that its total amount of output messages is considered relatively fewer.
	    For example, a pacemaker working for ten years outputs less than $2^{30}$ pairs of plaintext and ciphertext, which may lead to a relax of the total number of primitive rounds while retaining approximate the same security level \cite{chairforembeddedsecurity}.
	    \item {\it Limited Versatility--}The healthcare device where the encryption algorithm is to be implemented is usually function and operation focused, which makes encryption algorithms that have limited versatility rather welcome.
	\end{itemize}
	
	Considering the above implementation difficulties, security requirements, and feature preferences, our survey on lightweight symmetric encryption focuses on the algorithms that have a small block size or internal state, and can manage short keys. 
	Most of the candidate algorithms lie in block ciphers and stream ciphers due to the restrained resource. 
	For hash function based algorithm, only PHOTON \cite{guo2011photon} and Spongent \cite{bogdanov2011spongent} have ideally small internal state size.
	A summary of surveyed algorithms is showed in Table \ref{tbl:lwc}.
		
	\begin{table}
	\centering
    \caption{Comparison Among Suitable Lightweight Encryption Schemes}
    \label{tbl:lwc}
    \small
    \begin{tabular*}{27.5pc}{@{}|p{75pt}|p{20pt}|p{75pt}|p{55pt}|p{43pt}<{\raggedright}|@{}}
    %\begin{tabular*}{17.5pc}{@{}|p{35pt}|p{10pt}|p{45pt}|p{29pt}|p{28.5pt}<{\raggedright}|@{}}
    \hline
    \multicolumn{5}{@{}|p{16pc}|@{}}{Block Ciphers}\\\hline
    Name & Ref.& Key & Block & Rounds\\
    \hline
    Joltik&\cite{jean2015joltik}&64/80/96/128&64&24/32\\
    Mantis&\cite{beierle2016skinny}&128&64&14\\
    Skinny&\cite{beierle2016skinny}&64-384&64/128&32-56\\
    Qarma&\cite{avanzi2017qarma}&128/256&64/128&16/24\\
    T-TWINE&\cite{kubo2019tweakable}&80/128&64&36\\
    \hline
    \multicolumn{5}{@{}|p{16pc}|@{}}{Stream Ciphers}\\\hline
    Name & Ref.& Key & IV & IS\\
    \hline
    A2U2&\cite{david2011a2u2}&61&64&95\\
    Sprout&\cite{armknecht2015lightweight}&80&70&89\\
    Plantlet&\cite{mikhalev2016ciphers}&80&90&110\\
    \hline
    \multicolumn{5}{@{}|p{16pc}|@{}}{Hash}\\\hline
    Name & Ref.& Digest & Block & IS\\\hline
    PHOTON&\cite{guo2011photon}&80-256&16/32/64&100-288\\
    Spongent&\cite{bogdanov2011spongent}&80-256&8/16&88-272\\
\hline

\end{tabular*}
\end{table}

	\subsection{Discussion}
    For private data collected stored in many smart devices including wearable equipment that builds a personal data hub, we selectively surveyed privacy concerns on how the collected data can be accessed and how the sensitive data is transmitted by resource-restrained devices.
    Existing access control approaches help with the general purpose of controlling accessibility.
    However, most research works have not considered the usability that presents an essential requirement for personal scenarios.
    Another field in access control for future research is how to delicately assign accessibility according to the sensitivity of the collected data. 
    An example would be location information in residential areas, compared to public places, should be considered highly private and not suitable to be accessed by most applications.
	In terms of protection during transmission, lightweight encryption has showed practical promise in many IoT devices. 
	For healthcare devices, which could benefit from the seamless 5G network in the near future, characterised lightweight encryption schemes are expected to fit the challenging privacy scenarios.
	
	\section{Privacy in active data collection}\label{sec4}
	We now explore on privacy issues associated with the various uses of connected devices, e.g., to capture and record thescenery, people, activities, and other phenomena in the physical world. 
	
\begin{figure}
    \centerline{\includegraphics[width=28.5pc]{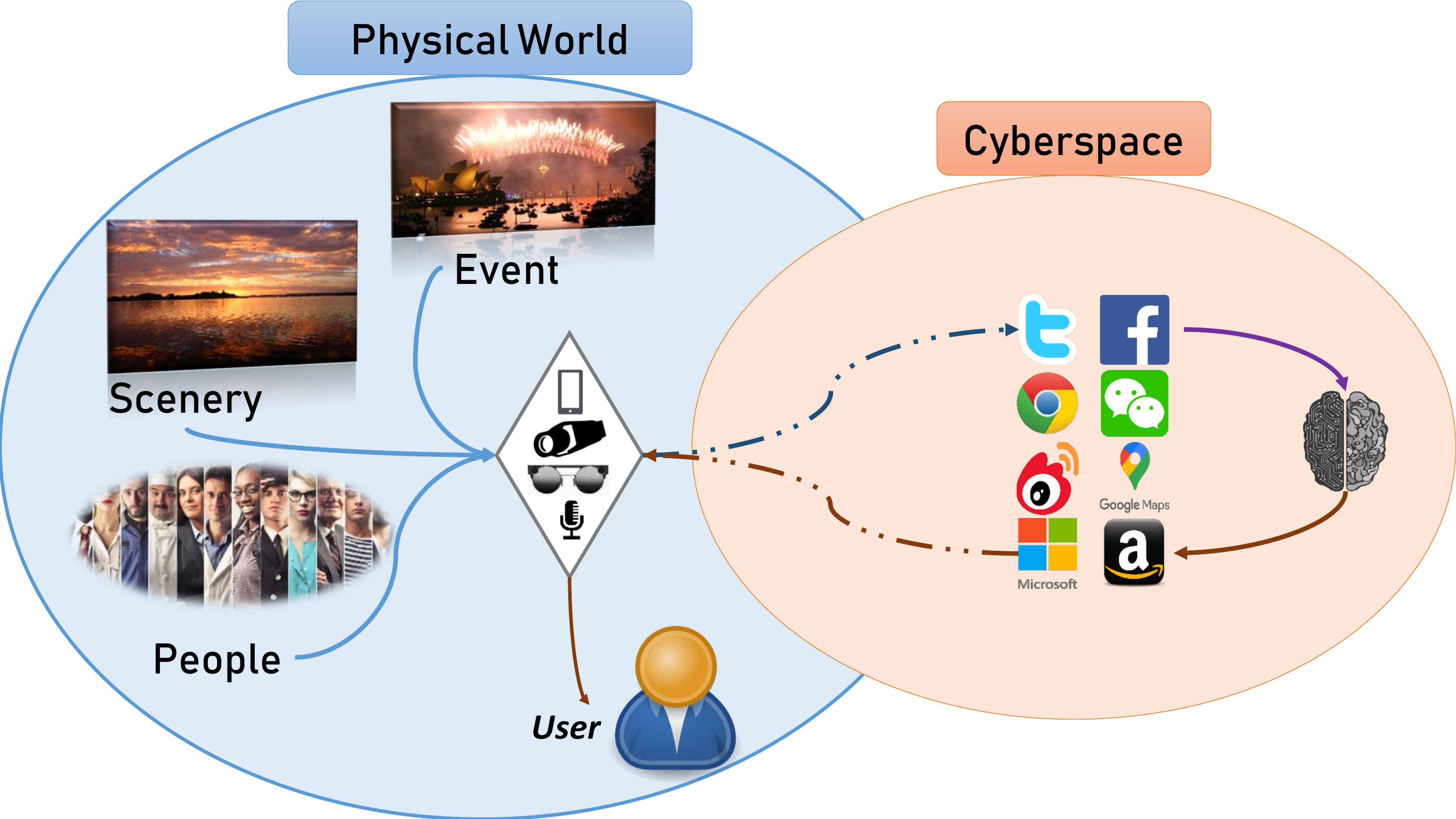}}
    \caption{Information leakage during active data collecting behaviours. People use smart devices collect and upload different types of data that simultaneously gets collected by cyberspace.}
    \label{fig:udc}
    \end{figure}

	\color{black}{People carry connected devices equipped with different types of sensors} and use them in their daily lives to retrieve information on a more and more frequent basis.
	Fig. \ref{fig:udc} illustrates the paradigm of how data collected via smart phones from the physical world may also be easily transmitted to different entities and services in cyberspace.
	In parallel with this active capturing and sharing, services have been developed based on a new environmental data collection model that individuals collect data on behalf of the service provider.
	These participators contribute their smart phones (or other smart devices) as sensors for applications helping collect environmental data such as surrounding noise, traffic conditions, thermal columns, etc..
	With an estimated 3 billion people having smart phones, this new model, known in the literature as \textit{participatory sensing} \cite{burke2006participatory,campbell2006people}, can accomplish large-scale sensing.
	%these two developing behaviours of ubiquitous data collecting share several same 
	
	\color{black}{However, personal information may get uploaded as media like photos and videos, which carry time and location information, or through being leaked from access by applications with permission.}
	Both these cases of data collecting via connected devices share common privacy concerns for user location information.
	At the same time, active data collecting encourages people to expand their collecting area to barren lands or distant villages in the mountains, valleys where network infrastructure is not available so that self-organising opportunistic networks need to be relied on.
	In this scenario, data collecting behaviours trigger the use of opportunistic networks and expose people's identities to untrusted authentications that raise identity privacy concerns, especially for participatory sensing where participants need records of sensing activities with identity for rewards.
	Below we take participatory sensing and opportunistic networks as two typical scenarios to survey related privacy concerns.
	
	\subsection{Participatory sensing and collector protection}
	Participatory sensing rushes onto the scene as a development of users' socialising data dissemination habits from users' activities on social media to remote data collection via multiple sensors on smart devices for services.
	When data is captured in social activities or participatory sensing, information about location, time and identity is simultaneously recorded and bound with the media data for uploading \cite{timberg_2012}.
	This bundle uploading happens at most user activities in cyberspace, from social media to ubiquitous participatory sensing.
	However, personal information that users think they have deleted adheres to its primary data in cyberspace \cite{cunningham2014next}, which then will be extracted, categorised, and analysed for long-term use.
	
	\subsubsection{Location privacy in participatory sensing}
% 	Protection of location privacy
    Location privacy is one of the most important concerns for participants in the participatory sensing activities because revealing location details within the sensing communication framework and to external entities may endanger personal privacy. 
    Nevertheless, the location data utility challenges the method used to protect location privacy, as low quality in location data can damage the significance of participatory sensing.
    Thus, the trade-off between location quality and location privacy has drawn attention from researchers.
    Many techniques have been studied and implemented for protecting location privacy while maintaining data utility. 
    Most of the techniques can be traced back to the following three influential methods:
    
    \begin{itemize}
        \item {\it Dummy locations--}The work proposed in \cite{kido2005anonymous}, brings the concept of dummy locations accompanying the true location of the user together in the query for services.
        %reveals the idea of dummy locations instead of actual locations in order to protect the participators or users privacy. 
        Their technique focuses on scenarios of an individual's location information for services. 
        Since the location information is from a user alone, the conventional technique of reducing accuracy \cite{gruteser2003anonymous} may not guarantee the user's privacy, while dummy locations make service providers unable to distinguish where the user truly is.
        \item {\it Obfuscation--}The work in \cite{duckham2005formal} proposed a novel approach of degrading location information with negotiation algorithms so that users privacy can be protected as location information obfuscated while the quality of service remains due to the negotiations.
        %oto protect the participators or users privacy by intentionally degrading the accuracy of the spatio-temporal information. 
        The obfuscation has since been developed as a major approach for location privacy; most of the following works have been carried out via the technique of perturbation \cite{ganti2008poolview} or generalisation \cite{ardagna2007location}. 
        \item {\it $k$-anonymity--}One of the most promising location privacy preserving approaches was originated from the $k$-anonymity concept \cite{sweeney2002k} by Gruteser and Grunwald \cite{gruteser2003anonymous}, which conceals the location information of a user within a group of $k\mbox{-}1$ other users.
        The $k$-anonymity approach is considered effective for location privacy in general scenarios where the other $k\mbox{-}1$ users are not difficult to find.
        However, it does not perform well on continuous location information transmission and sometimes has too low accuracy resulting in service failure \cite{mokbel2006new}.
    \end{itemize}
     
   	\subsection{Privacy in opportunistic networks}
	When people travel between cities, states or rural regions, they need to connect to different networks for services.	%PEOPLE WANT TO CONNECT TO KEEP BEING SERVICED
	A surfacing type of network, named opportunistic networks (OppNets), plays an important role in maintaining the connections in a simple, easy, and convenient way.
	Opportunistic networks usually are built on convenient communications technologies for short-term connectivity when there is limited or no existing traditional network infrastructure.
	Although these networks may only provide connections with limited bandwidth, high latency and packet-loss rate, applications still upload collected data.
	Since opportunistic networks are provided as a necessary alternative, their nature of expanding from node to node makes them self-organising, which unfortunately come with some undesired features. 
	OppNets use store-carry and forward mechanism to connect and extend the network because the path between the source and destination does not exist
	\cite{kumar2017authentication,tsai2016provably,irshad2018cryptanalysis}.
	For messages or packets through intermediate nodes, any of these intermediate nodes can turn malicious and exploit received information.
	Thus, countermeasures are needed to protect user privacy considering the forwarded messages may contain personal information, especially during the authentication procedure.

	\subsubsection{Anonymous authentication for OppNets}
	%anonymous comm authen
	The node authentication procedure plays an essential role in OppNets: it confirms a node's credentials and prevents unauthorised nodes from joining/accessing the network; it also authenticates the integrity of the packet received by nodes.
	As mentioned above, the path between two distant nodes in OppNets does not exist, so it is a challenging task for researchers to construct efficient authentication algorithms and solutions validating connectivity between nodes. 
	\color{black}{It adds further difficulties for authentication algorithms to achieve anonymity in a dynamic OppNets environment}.
	So far in the literature, only a limited range of anonymous authentication algorithms succeed in OppNets. These are compared in Table \ref{tbl:aao}.
	
	In 2012, Carver and Lin utilised group-oriented broadcast encryption based on pairing and identity-based signature, and constructed an authentication scheme for OppNets based on Bluetooth and 3G communications \cite{carver2012privacy}.
	In their work, the forwarding packet does not require knowledge of recipients, so the privacy of users is partially protected. The sender's information, however, will be acknowledged after authentication.
	In addition, their work relies on a trusted third party for key generation and group categorisation, which leads to a critical risk on system security and user privacy if the third-party is compromised \cite{avoussoukpo2020ensuring}.
	
	In 2015, Guo et al. proposed an authentication protocol for OppNets that fully protects user privacy \cite{guo2015authenticating}.
	The authentication is designed for short-term and limited-time wireless network environment where node registration is performed by a selected super node \cite{guo2015authenticating}.
	In their protocol, both symmetric and asymmetric encryption are adopted as countermeasures for security threats and attacks while privacy is achieved by protecting user identities with hash functions.
	Later in 2017, Kumar et al. proposed a security algorithm based on \cite{guo2015authenticating} that fulfils the authentication requirements in OppNets and protects user privacy.
	Different from \cite{guo2015authenticating}, Kumar et al.'s work utilises dynamic user identities for the key exchange mechanism and RSA encryption for message transmission, so both user and data privacy can be protected.
	
	Besides the aforementioned works that heavily rely on symmetric and asymmetric encryption, Kuo et al. proposed an efficient and secure anonymous authentication scheme utilising hash functionality and point operations \cite{kuo2014efficient}.
	Although their work was not motivated for OppNets, the roaming authentication feature can be applied to OppNets environments with enhanced performance and additional security properties.
	
	\begin{table}
	\centering
\caption{Comparison Among Anonymous Authentication for OppNets}
\label{tbl:aao}
\small
\begin{tabular*}{27.5pc}{@{}|p{20pt}|p{183pt}<{\raggedright}|p{90pt}<{\raggedright}|@{}}
\hline
Ref.& 
Technique& 
Feature \\
\hline
\cite{carver2012privacy}&Broadcast encryption&Partial privacy\\
\multirow{2}{*}{\cite{guo2015authenticating}}&\multirow{2}{183pt}{Symmetric \& asymmetric encryption}&\multirow{2}{*}{Hashed user ID}\\
&&\\
\cite{kumar2017authentication}&RSA encryption&Dynamic user ID\\
\hline
\multirow{2}{*}{\cite{kuo2014efficient}}&\multirow{2}{150pt}{Hash functionality\\Point operation}&\multirow{2}{*}{Encryption free}\\
&&\\

\hline

\end{tabular*}
\end{table}
	
	\subsection{Discussion}
	The existence of users and data generated by users raises concerns about protecting user privacy during active data collecting activities.
	From these concerns, we survey the problems of location privacy in participatory sensing and anonymous authentication in OppNets.
	To protect a participant's location data, a compromise in the quality of location data is usually the trade-off, although many efforts have been made to mitigate the effect.
	Concerning identity privacy in OppNets, existing solutions heavily rely on encryption techniques, which can be expensive considering heterogeneous devices.
	Encryption free anonymous authentication requires more research as it potentially has more application scenarios.
	
	\section{Future vision on complex privacy problems}\label{sec5}
	There are many complex privacy problems already identified that soon could have a considerable impact on industry as well as our daily lives. We deliver our future vision on three of these privacy concerns, i.e., trajectory privacy, privacy in smart metering, and involuntary information leakage with ambient intelligence.
	
	\subsection{Trajectory privacy}
	When we apply the traditional scenario of cookie privacy concerns to mobile applications, users' trajectories become at risk due to location information embedded in cookie logs.
	Cookie logs in cyberspace may contain high-quality user location information, which can be collected directly using GPS coordinates with a user's fast consent to an unexplained location service permission requirement, or indirectly collected with location tags from a local network or service provider in the physical world.
	This potential privacy breach should be categorised to a more dangerous level than web browsing history or personal preference logs.
	More detailed physical activities, routine habits or even mental status can be inferred by analytical work on user trajectories.
	The infamous Uber travel history leakage lawsuit in 2017 \cite{braun_2017} is a relevant example.
	\cite{wang2017fingerprint} developed a privacy analysis system on user login records and physical context information, and deepened the understanding of user physical-world privacy leakage via cyberspace privacy leakage.
	It becomes clear that user trajectories can be discovered and confirmed when third parties analyse their cookie logs as users move and browse in their daily lives let alone potential exogenous records of GPS coordinates.
	These cookie logs may further be exchanged with other analytics companies for centralised analysis connecting with other web activities, exposing exposes private physical trajectory to more entities.
	Compared with other private data, physical trajectory is more effective on re-identification with auditing relevant activity logs at locations and comparing differential timelines.
	The balance between utility and privacy with location data has drawn much research attention. 
	However, for this physical world trajectory leakage via user cyberspace data, further research efforts are required.

	\subsection{Privacy in smart metering}
	Another emerging privacy concern is in smart metering in smart energy supply networks, which is considered as the next evolutionary step in the industry\cite{saxena2015state}.
	For smart energy supply, smart metering helps evaluate the status of a smart energy grid and contribute detailed usage information for efficient resource distribution.
	To achieve this, many smart meters and sensors are needed between consuming points and monitoring centres \cite{wu2016bi2g,wu2016big2,uribe2016state} as well as the networks used for sending and receiving precise metering data.
	Thus, privacy concerns arise as the consumption usage data are transmitted and stored in plaintext.

	Since the consumption usage data include private activities of consumers, this information, which is invaluable to service providers \cite{kumar2019smart}, together with identity/location tags and physical address information, forms a large attack surface drawing adversaries' attention.

	For this privacy concern, a key obstacle is the limited resource the smart meters have to perform strong cryptography \cite{kumar2019smart}.
	Therefore, it remains a challenge for future research focusing on cryptography-based mechanisms that must provide confidentiality, while minimising resource consumption.

	\subsection{Involuntary privacy leakage with ambient intelligence} 
	Ambient intelligence makes environments sensitive to users as sensors deliver the change of the state of an environment for computing faster and better services to a user \cite{cook2009ambient}.
	As smart devices play more important roles in our daily lives, a mobile ambient intelligence is subtly forming with different functionalities.
	Beneficial services and better user experience provided by smart phones as well as other wearable gadgets encourage users to give the green light to data collection from smart devices, which implicitly offers a convenient way for private information to be collected and leaked.
	Since these smart devices are connected to the internet acting as the gate between the physical world and cyberspace, potential attacks on collected private data may come from both dimensions.
	An example of an attack from cyberspace can be seen from \cite{judd_2020} that reported an incident of smart watches leaking its users' real-time location data, which is described as ``pretty common" by Manuel.
	Although attacks from cyberspace focusing on software bugs can be fixed in hours, privacy breaches can occur due to incomplete test procedures, even if there are many effective security approaches in place \cite{judd_2020}.
	Now we consider potential attacks from the physical world, which can be unintentional or malicious.
	For unintentional attacks, an example would be a stranger's face being captured accidentally by other people in an event, who then upload the photo to social media networks.
	Although the owner of the photo does not know who the stranger is, machine learning algorithms behind social media can potentially identify the stranger. This information would include details of the time the photo was taken as well as where the photo was taken.
	With more and more resources that smart phones and other smart devices are equipped with, people can now capture and record almost every detail of what they see, hear and experience in their daily lives.
	So the fate of information about those captured also lies in their hands.
	This leads to two questions for future research: would smart devices be able to realise what kind of environment they are in and apply the correct privacy strategy, and could smart devices notify their owners about what kind of environment they are currently in and warn them to behave with caution.

	\section{Conclusion}\label{sec6}
	User data is being increasingly collected by new pervasive technologies for further data analytics in cyberspace, which raises many privacy concerns.
	In this paper, we comprehensively surveyed and studied these concerns that are classified as data exhaust tracing, personal data hub and active data collection.
	%how user data is collected and the relation between user activities and cyberspace.
	Among them, six privacy concerns including identity tracking from browsing exhaust, security in delicate transmission, and privacy in opportunistic networks, are featured and defined with a real-world application scenario.
	The privacy concerns are studied in the scenarios for their properties and the differences compared to similar but well-known problems are highlighted.
	With these highlighted differences, we surveyed existing technologies for countermeasures, compared their methodologies and performance, and discussed implementation difficulties and research gaps.
	With these impending and overhanging concerns analysed, we further analysed the privacy in complex problems of user trajectory, smart metering and ambient intelligence.
	In these complex problems, privacy concerns involve elements from many aspects, which present many challenges with complicated difficulties.
	Possible research suggestions and further directions are discussed.

    \section*{Acknowledgement}
    The work has been supported by the Cyber Security Research Centre Limited whose activities are partially funded by the Australian Government’s Cooperative Research Centres Programme.




\bibliographystyle{elsarticle-num}
\bibliography{reference_bib}

\begin{thebibliography}{10}
\expandafter\ifx\csname url\endcsname\relax
  \def\url#1{\texttt{#1}}\fi
\expandafter\ifx\csname urlprefix\endcsname\relax\def\urlprefix{URL }\fi
\expandafter\ifx\csname href\endcsname\relax
  \def\href#1#2{#2} \def\path#1{#1}\fi

\bibitem{conti2017internet}
M.~Conti, A.~Passarella, S.~K. Das, The internet of people (iop): A new wave in
  pervasive mobile computing, Pervasive and Mobile Computing 41 (2017) 1--27.

\bibitem{westin1997whatever}
A.~F. Westin, Whatever works": The american public's attitudes toward
  regulation and self-regulation on consumer privacy issues, US Dep't Of
  Commerce, Privacy And Self-Regulation In The Information Age (1997).

\bibitem{zawadzinski_sweeney_2019}
M.~Zawadziński, M.~Sweeney,
  \href{https://clearcode.cc/blog/adtech-id-problem/}{Identity in adtech:
  Unravelling the id problem - clearcode blog} (Sep 2019).
\newline\urlprefix\url{https://clearcode.cc/blog/adtech-id-problem/}

\bibitem{soltani2010flash}
A.~Soltani, S.~Canty, Q.~Mayo, L.~Thomas, C.~J. Hoofnagle, Flash cookies and
  privacy, in: 2010 AAAI Spring Symposium Series, 2010.

\bibitem{eckersley2010unique}
P.~Eckersley, How unique is your web browser?, in: International Symposium on
  Privacy Enhancing Technologies Symposium, Springer, 2010, pp. 1--18.

\bibitem{acar2014web}
G.~Acar, C.~Eubank, S.~Englehardt, M.~Juarez, A.~Narayanan, C.~Diaz, The web
  never forgets: Persistent tracking mechanisms in the wild, in: Proceedings of
  the 2014 ACM SIGSAC Conference on Computer and Communications Security, 2014,
  pp. 674--689.

\bibitem{ravichandran_korula_2019}
D.~Ravichandran, N.~Korula,
  \href{https://services.google.com/fh/files/misc/disabling\_third-party\_cookies\_publisher\_revenue.pdf}{Effect
  of disabling third-party cookies on publisher revenue} (Aug 2019).
\newline\urlprefix\url{https://services.google.com/fh/files/misc/disabling\_third-party\_cookies\_publisher\_revenue.pdf}

\bibitem{merzdovnik2017block}
G.~Merzdovnik, M.~Huber, D.~Buhov, N.~Nikiforakis, S.~Neuner, M.~Schmiedecker,
  E.~Weippl, Block me if you can: A large-scale study of tracker-blocking
  tools, in: 2017 IEEE European Symposium on Security and Privacy (EuroS\&P),
  IEEE, 2017, pp. 319--333.

\bibitem{ad-blocker}
Ad-Blocker, \href{http://www.ad-blocker.org/}{Block annoying ads to surf web
  faster}.
\newline\urlprefix\url{http://www.ad-blocker.org/}

\bibitem{adblock}
AdBlock, \href{https://getadblock.com/}{Surf the web without annoying pop ups
  and ads!}
\newline\urlprefix\url{https://getadblock.com/}

\bibitem{adblock-plus}
A.~Plus, \href{https://adblockplus.org/}{Adblock plus: The world's no. 1 free
  ad blocker}.
\newline\urlprefix\url{https://adblockplus.org/}

\bibitem{hill_2020}
R.~Hill, \href{https://github.com/gorhill/uBlock}{gorhill/ublock} (Mar 2020).
\newline\urlprefix\url{https://github.com/gorhill/uBlock}

\bibitem{easylist}
EasyList, \href{https://easylist.to/}{Overview}.
\newline\urlprefix\url{https://easylist.to/}

\bibitem{ghostery}
Ghostery, \href{https://www.ghostery.com/}{Ghostery makes the web cleaner,
  faster and safer!}
\newline\urlprefix\url{https://www.ghostery.com/}

\bibitem{disconnect}
Disconnect, \href{https://disconnect.me/}{Take back your privacy}.
\newline\urlprefix\url{https://disconnect.me/}

\bibitem{abineblur}
I.~Abine, \href{https://dnt.abine.com/\#help/faq/faq-unblocked}{Keep your web
  activity and personal info private}.
\newline\urlprefix\url{https://dnt.abine.com/\#help/faq/faq-unblocked}

\bibitem{ghosteryins}
Ghostery, \href{https://www.ghostery.com/insights/}{Insights}.
\newline\urlprefix\url{https://www.ghostery.com/insights/}

\bibitem{disconnectfq}
Disconnect, \href{shorturl.at/dnxZ2}{Faq}.
\newline\urlprefix\url{shorturl.at/dnxZ2}

\bibitem{abineincbp}
I.~Abine, \href{https://dnt.abine.com/\#premiumreg}{Keep your web activity and
  personal info private}.
\newline\urlprefix\url{https://dnt.abine.com/\#premiumreg}

\bibitem{eff_2018}
EFF, \href{https://www.eff.org/privacybadger}{Privacy badger} (Oct 2018).
\newline\urlprefix\url{https://www.eff.org/privacybadger}

\bibitem{eff_2019}
EFF,
  \href{https://www.eff.org/privacybadger/faq\#What-is-Privacy-Badger}{Privacy
  badger} (Jul 2019).
\newline\urlprefix\url{https://www.eff.org/privacybadger/faq\#What-is-Privacy-Badger}

\bibitem{storey2017future}
G.~Storey, D.~Reisman, J.~Mayer, A.~Narayanan, The future of ad blocking: An
  analytical framework and new techniques, arXiv preprint arXiv:1705.08568
  (2017).

\bibitem{tigas2019percival}
P.~Tigas, S.~T. King, B.~Livshits, et~al., Percival: Making in-browser
  perceptual ad blocking practical with deep learning, arXiv preprint
  arXiv:1905.07444 (2019).

\bibitem{paraska_2018}
O.~Paraska, \href{shorturl.at/imrxM}{Towards more intelligent ad blocking on
  the web} (Jun 2018).
\newline\urlprefix\url{shorturl.at/imrxM}

\bibitem{tramer2019adversarial}
F.~Tram{\`e}r, P.~Dupr{\'e}, G.~Rusak, G.~Pellegrino, D.~Boneh, Adversarial:
  Perceptual ad blocking meets adversarial machine learning, in: Proceedings of
  the 2019 ACM SIGSAC Conference on Computer and Communications Security, 2019,
  pp. 2005--2021.

\bibitem{storey2017adh}
G.~Storey, D.~Reisman, J.~Mayer, A.~Narayanan,
  \href{https://chrome.google.com/webstore/detail/perceptual-ad-highlighter/mahgiflleahghaapkboihnbhdplhnchp}{Perceptual
  ad highlighter}.
\newline\urlprefix\url{https://chrome.google.com/webstore/detail/perceptual-ad-highlighter/mahgiflleahghaapkboihnbhdplhnchp}

\bibitem{adblockplusst}
AdblockPlus, \href{https://adblock.ai/}{Developed by adblockplus}.
\newline\urlprefix\url{https://adblock.ai/}

\bibitem{iqbal2020adgraph}
U.~Iqbal, P.~Snyder, S.~Zhu, B.~Livshits, Z.~Qian, Z.~Shafiq, Adgraph: A
  graph-based approach to ad and tracker blocking, in: Proc. of IEEE Symposium
  on Security and Privacy, 2020.

\bibitem{dwork2006calibrating}
C.~Dwork, F.~McSherry, K.~Nissim, A.~Smith, Calibrating noise to sensitivity in
  private data analysis, in: Theory of cryptography conference, Springer, 2006,
  pp. 265--284.

\bibitem{yang2012differential}
Y.~Yang, Z.~Zhang, G.~Miklau, M.~Winslett, X.~Xiao, Differential privacy in
  data publication and analysis, in: Proceedings of the 2012 ACM SIGMOD
  International Conference on Management of Data, ACM, 2012, pp. 601--606.

\bibitem{35dwork2011differential}
C.~Dwork, Differential privacy, Encyclopedia of Cryptography and Security
  (2011) 338--340.

\bibitem{erlingsson2014rappor}
{\'U}.~Erlingsson, V.~Pihur, A.~Korolova, Rappor: Randomized aggregatable
  privacy-preserving ordinal response, in: Proceedings of the 2014 ACM SIGSAC
  conference on computer and communications security, 2014, pp. 1054--1067.

\bibitem{apple_2017}
D.~P.~T. Apple,
  \href{https://machinelearning.apple.com/2017/12/06/learning-with-privacy-at-scale.html}{Learning
  with privacy at scale} (Dec 2017).
\newline\urlprefix\url{https://machinelearning.apple.com/2017/12/06/learning-with-privacy-at-scale.html}

\bibitem{warner1965randomized}
S.~L. Warner, Randomized response: A survey technique for eliminating evasive
  answer bias, Journal of the American Statistical Association 60~(309) (1965)
  63--69.

\bibitem{bloom1970space}
B.~H. Bloom, \href{http://doi.acm.org/10.1145/362686.362692}{Space/time
  trade-offs in hash coding with allowable errors}, Commun. ACM 13~(7) (1970)
  422--426.
\newblock \href {https://doi.org/10.1145/362686.362692}
  {\path{doi:10.1145/362686.362692}}.
\newline\urlprefix\url{http://doi.acm.org/10.1145/362686.362692}

\bibitem{hsu2012distributed}
J.~Hsu, S.~Khanna, A.~Roth, Distributed private heavy hitters, in:
  International Colloquium on Automata, Languages, and Programming, Springer,
  2012, pp. 461--472.

\bibitem{bassily2017practical}
R.~Bassily, K.~Nissim, U.~Stemmer, A.~G. Thakurta, Practical locally private
  heavy hitters, in: Advances in Neural Information Processing Systems, 2017,
  pp. 2288--2296.

\bibitem{wang2019locally}
T.~Wang, N.~Li, S.~Jha, Locally differentially private heavy hitter
  identification, IEEE Transactions on Dependable and Secure Computing (2019).

\bibitem{wang2017locally}
T.~Wang, J.~Blocki, N.~Li, S.~Jha, Locally differentially private protocols for
  frequency estimation, in: 26th $\{$USENIX$\}$ Security Symposium
  ($\{$USENIX$\}$ Security 17), 2017, pp. 729--745.

\bibitem{wang2018locally}
T.~Wang, N.~Li, S.~Jha, Locally differentially private frequent itemset mining,
  in: 2018 IEEE Symposium on Security and Privacy (SP), IEEE, 2018, pp.
  127--143.

\bibitem{cormode2018marginal}
G.~Cormode, T.~Kulkarni, D.~Srivastava, Marginal release under local
  differential privacy, in: Proceedings of the 2018 International Conference on
  Management of Data, 2018, pp. 131--146.

\bibitem{ding2017collecting}
B.~Ding, J.~Kulkarni, S.~Yekhanin, Collecting telemetry data privately, in:
  Advances in Neural Information Processing Systems, 2017, pp. 3571--3580.

\bibitem{qin2017generating}
Z.~Qin, T.~Yu, Y.~Yang, I.~Khalil, X.~Xiao, K.~Ren, Generating synthetic
  decentralized social graphs with local differential privacy, in: Proceedings
  of the 2017 ACM SIGSAC Conference on Computer and Communications Security,
  2017, pp. 425--438.

\bibitem{mcmahan2017learning}
H.~B. McMahan, D.~Ramage, K.~Talwar, L.~Zhang, Learning differentially private
  language models without losing accuracy, arXiv preprint arXiv:1710.06963
  (2017).

\bibitem{nguyen2016collecting}
T.~T. Nguy{\^e}n, X.~Xiao, Y.~Yang, S.~C. Hui, H.~Shin, J.~Shin, Collecting and
  analyzing data from smart device users with local differential privacy, arXiv
  preprint arXiv:1606.05053 (2016).

\bibitem{jia2019calibrate}
J.~Jia, N.~Z. Gong, Calibrate: Frequency estimation and heavy hitter
  identification with local differential privacy via incorporating prior
  knowledge, in: IEEE INFOCOM 2019-IEEE Conference on Computer Communications,
  IEEE, 2019, pp. 2008--2016.

\bibitem{garcia2010modular}
O.~Garcia-Morchon, K.~Wehrle, Modular context-aware access control for medical
  sensor networks, in: Proceedings of the 15th ACM symposium on Access control
  models and technologies, 2010, pp. 129--138.

\bibitem{ray2017using}
I.~Ray, B.~Alangot, S.~Nair, K.~Achuthan, Using attribute-based access control
  for remote healthcare monitoring, in: 2017 Fourth International Conference on
  Software Defined Systems (SDS), IEEE, 2017, pp. 137--142.

\bibitem{118DBLP:conf/icws/SalamaYWPB17}
U.~Salama, L.~Yao, X.~Wang, H.~Paik, A.~Beheshti,
  \href{https://doi.org/10.1109/ICWS.2017.111}{Multi-level privacy-preserving
  access control as a service for personal healthcare monitoring}, in:
  I.~Altintas, S.~Chen (Eds.), 2017 {IEEE} International Conference on Web
  Services, {ICWS} 2017, Honolulu, HI, USA, June 25-30, 2017, {IEEE}, 2017, pp.
  878--881.
\newblock \href {https://doi.org/10.1109/ICWS.2017.111}
  {\path{doi:10.1109/ICWS.2017.111}}.
\newline\urlprefix\url{https://doi.org/10.1109/ICWS.2017.111}

\bibitem{ravidas2019access}
S.~Ravidas, A.~Lekidis, F.~Paci, N.~Zannone, Access control in
  internet-of-things: A survey, Journal of Network and Computer Applications
  144 (2019) 79--101.

\bibitem{kim2011access}
T.~H.-J. Kim, L.~Bauer, J.~Newsome, A.~Perrig, J.~Walker, Access right
  assignment mechanisms for secure home networks, Journal of Communications and
  Networks 13~(2) (2011) 175--186.

\bibitem{tian2017smartauth}
Y.~Tian, N.~Zhang, Y.-H. Lin, X.~Wang, B.~Ur, X.~Guo, P.~Tague, Smartauth:
  User-centered authorization for the internet of things, in: 26th
  $\{$USENIX$\}$ Security Symposium ($\{$USENIX$\}$ Security 17), 2017, pp.
  361--378.

\bibitem{alshehri2013secure}
S.~Alshehri, R.~K. Raj, Secure access control for health information sharing
  systems, in: 2013 IEEE international conference on healthcare informatics,
  IEEE, 2013, pp. 277--286.

\bibitem{burnap2012protecting}
P.~R. Burnap, I.~Spasi{\'c}, W.~A. Gray, J.~C. Hilton, O.~F. Rana, G.~Elwyn,
  Protecting patient privacy in distributed collaborative healthcare
  environments by retaining access control of shared information, in: 2012
  International Conference on Collaboration Technologies and Systems (CTS),
  IEEE, 2012, pp. 490--497.

\bibitem{heydari2019towards}
M.~Heydari, A.~Mylonas, V.~Katos, D.~Gritzalis, Towards indeterminacy-tolerant
  access control in iot, in: Handbook of Big Data and IoT Security, Springer,
  2019, pp. 53--71.

\bibitem{42rahman2017privacy}
F.~Rahman, M.~Z.~A. Bhuiyan, S.~I. Ahamed, A privacy preserving framework for
  rfid based healthcare systems, Future generation computer systems 72 (2017)
  339--352.

\bibitem{diez2019lightweight}
F.~P. Diez, D.~S. Touceda, J.~M.~S. C{\'a}mara, S.~Zeadally, Lightweight access
  control system for wearable devices, IT Professional 21~(1) (2019) 50--58.

\bibitem{Centeno2018performance}
J.~K.~M. {Centeno}, P.~S. {Chhabra}, C.~L. {Fianza}, I.~{Montes-Austria},
  R.~{Ocampo}, Performance analysis of encryption algorithms on smartwatches,
  in: TENCON 2018 - 2018 IEEE Region 10 Conference, 2018, pp. 0162--0166.

\bibitem{biryukov2017state}
A.~Biryukov, L.~P. Perrin, State of the art in lightweight symmetric
  cryptography (2017).

\bibitem{masoud2015power}
M.~Masoud, I.~Jannoud, A.~Ahmad, H.~Al-Shobaky, The power consumption cost of
  data encryption in smartphones, in: 2015 International Conference on Open
  Source Software Computing (OSSCOM), IEEE, 2015, pp. 1--6.

\bibitem{ronen2017iot}
E.~Ronen, A.~Shamir, A.-O. Weingarten, C.~O’Flynn, Iot goes nuclear: Creating
  a zigbee chain reaction, in: 2017 IEEE Symposium on Security and Privacy
  (SP), IEEE, 2017, pp. 195--212.

\bibitem{rob16}
M.~Robshaw,
  \href{https://www.nist.gov/system/files/documents/2016/10/17/robshaw-presentation-lwc2016.pdf}{Lightweight
  cryptography and rain rfid}.
\newline\urlprefix\url{https://www.nist.gov/system/files/documents/2016/10/17/robshaw-presentation-lwc2016.pdf}

\bibitem{chairforembeddedsecurity}
G.~Leander, V.~Nikov, C.~Rechberger, V.~Rijmen,
  \href{https://www.emsec.ruhr-uni-bochum.de/research/research\_startseite/prince-challenge/}{The
  prince challenge}.
\newline\urlprefix\url{https://www.emsec.ruhr-uni-bochum.de/research/research\_startseite/prince-challenge/}

\bibitem{guo2011photon}
J.~Guo, T.~Peyrin, A.~Poschmann, The photon family of lightweight hash
  functions, in: Annual Cryptology Conference, Springer, 2011, pp. 222--239.

\bibitem{bogdanov2011spongent}
A.~Bogdanov, M.~Kne{\v{z}}evi{\'c}, G.~Leander, D.~Toz, K.~Var{\i}c{\i},
  I.~Verbauwhede, Spongent: A lightweight hash function, in: International
  Workshop on Cryptographic Hardware and Embedded Systems, Springer, 2011, pp.
  312--325.

\bibitem{jean2015joltik}
J.~Jean, I.~Nikoli{\'c}, T.~Peyrin, Joltik v1. 3, CAESAR Round 2 (2015).

\bibitem{beierle2016skinny}
C.~Beierle, J.~Jean, S.~K{\"o}lbl, G.~Leander, A.~Moradi, T.~Peyrin, Y.~Sasaki,
  P.~Sasdrich, S.~M. Sim, The skinny family of block ciphers and its
  low-latency variant mantis, in: Annual International Cryptology Conference,
  Springer, 2016, pp. 123--153.

\bibitem{avanzi2017qarma}
R.~Avanzi, The qarma block cipher family. almost mds matrices over rings with
  zero divisors, nearly symmetric even-mansour constructions with
  non-involutory central rounds, and search heuristics for low-latency s-boxes,
  IACR Transactions on Symmetric Cryptology (2017) 4--44.

\bibitem{kubo2019tweakable}
H.~Kubo, Y.~Funabiki, A.~Bogdanov, S.~Morioka, T.~Isobe, Tweakable twine:
  Building a tweakable block cipher on generalized feistel structure, in:
  Advances in Information and Computer Security: 14th International Workshop on
  Security, IWSEC 2019, Tokyo, Japan, August 28-30, 2019, Proceedings, Vol.
  11689, Springer, 2019, p. 129.

\bibitem{david2011a2u2}
M.~David, D.~C. Ranasinghe, T.~Larsen, A2u2: a stream cipher for printed
  electronics rfid tags, in: 2011 IEEE International Conference on RFID, IEEE,
  2011, pp. 176--183.

\bibitem{armknecht2015lightweight}
F.~Armknecht, V.~Mikhalev, On lightweight stream ciphers with shorter internal
  states, in: International Workshop on Fast Software Encryption, Springer,
  2015, pp. 451--470.

\bibitem{mikhalev2016ciphers}
V.~Mikhalev, F.~Armknecht, C.~M{\"u}ller, On ciphers that continuously access
  the non-volatile key, IACR Transactions on Symmetric Cryptology (2016)
  52--79.

\bibitem{burke2006participatory}
J.~A. Burke, D.~Estrin, M.~Hansen, A.~Parker, N.~Ramanathan, S.~Reddy, M.~B.
  Srivastava, Participatory sensing (2006).

\bibitem{campbell2006people}
A.~T. Campbell, S.~B. Eisenman, N.~D. Lane, E.~Miluzzo, R.~A. Peterson,
  People-centric urban sensing, in: Proceedings of the 2nd annual international
  workshop on Wireless internet, 2006, pp. 18--es.

\bibitem{timberg_2012}
C.~Timberg,
  \href{https://www.independent.co.uk/news/world/europe/austrian-student-challenges-facebooks-use-of-personal-data-8219155.html}{Austrian
  student challenges facebook's use of personal data} (Oct 2012).
\newline\urlprefix\url{https://www.independent.co.uk/news/world/europe/austrian-student-challenges-facebooks-use-of-personal-data-8219155.html}

\bibitem{cunningham2014next}
M.~Cunningham, Next generation privacy: The internet of things, data exhaust,
  and reforming regulation by risk of harm, Groningen Journal of International
  Law 2 (2014).

\bibitem{kido2005anonymous}
H.~Kido, Y.~Yanagisawa, T.~Satoh, An anonymous communication technique using
  dummies for location-based services, in: ICPS'05. Proceedings. International
  Conference on Pervasive Services, 2005., IEEE, 2005, pp. 88--97.

\bibitem{gruteser2003anonymous}
M.~Gruteser, D.~Grunwald, Anonymous usage of location-based services through
  spatial and temporal cloaking, in: Proceedings of the 1st international
  conference on Mobile systems, applications and services, 2003, pp. 31--42.

\bibitem{duckham2005formal}
M.~Duckham, L.~Kulik, A formal model of obfuscation and negotiation for
  location privacy, in: International conference on pervasive computing,
  Springer, 2005, pp. 152--170.

\bibitem{ganti2008poolview}
R.~K. Ganti, N.~Pham, Y.-E. Tsai, T.~F. Abdelzaher, Poolview: stream privacy
  for grassroots participatory sensing, in: Proceedings of the 6th ACM
  conference on Embedded network sensor systems, 2008, pp. 281--294.

\bibitem{ardagna2007location}
C.~A. Ardagna, M.~Cremonini, E.~Damiani, S.~D.~C. Di~Vimercati, P.~Samarati,
  Location privacy protection through obfuscation-based techniques, in: IFIP
  Annual Conference on Data and Applications Security and Privacy, Springer,
  2007, pp. 47--60.

\bibitem{sweeney2002k}
L.~Sweeney, k-anonymity: A model for protecting privacy, International Journal
  of Uncertainty, Fuzziness and Knowledge-Based Systems 10~(05) (2002)
  557--570.

\bibitem{mokbel2006new}
M.~F. Mokbel, C.-Y. Chow, W.~G. Aref, The new casper: Query processing for
  location services without compromising privacy, in: Proceedings of the 32nd
  international conference on Very large data bases, 2006, pp. 763--774.

\bibitem{kumar2017authentication}
P.~Kumar, N.~Chauhan, N.~Chand, Authentication with privacy preservation in
  opportunistic networks, in: 2017 International Conference on Inventive
  Communication and Computational Technologies (ICICCT), IEEE, 2017, pp.
  183--188.

\bibitem{tsai2016provably}
J.-L. Tsai, N.-W. Lo, Provably secure anonymous authentication with batch
  verification for mobile roaming services, Ad Hoc Networks 44 (2016) 19--31.

\bibitem{irshad2018cryptanalysis}
A.~Irshad, M.~Sher, B.~A. Alzahrani, A.~Albeshri, S.~A. Chaudhry, S.~Kumari,
  Cryptanalysis and improvement of a multi-server authentication protocol by lu
  et al., KSII Transactions on Internet \& Information Systems 12~(1) (2018).

\bibitem{carver2012privacy}
C.~Carver, X.~Lin, A privacy-preserving proximity friend notification scheme
  with opportunistic networking, in: 2012 IEEE International Conference on
  Communications (ICC), IEEE, 2012, pp. 5387--5392.

\bibitem{avoussoukpo2020ensuring}
C.~B. Avoussoukpo, C.~Xu, M.~Tchenagnon, Ensuring users privacy and mutual
  authentication in opportunistic networks: A survey, International Journal of
  Network Security 22~(1) (2020) 118--125.

\bibitem{guo2015authenticating}
{Ming-Huang Guo}, {Horng-Twu Liaw}, {Meng-Yu Chiu}, {Li-Ping Tsai},
  Authenticating with privacy protection in opportunistic networks, in: 2015
  11th International Conference on Heterogeneous Networking for Quality,
  Reliability, Security and Robustness (QSHINE), 2015, pp. 375--380.

\bibitem{kuo2014efficient}
W.-C. Kuo, H.-J. Wei, J.-C. Cheng, An efficient and secure anonymous mobility
  network authentication scheme, journal of information security and
  applications 19~(1) (2014) 18--24.

\bibitem{braun_2017}
E.~Braun, \href{shorturl.at/CEJM9}{Un français demande 45 millions d'euros à
  uber pour avoir précipité son divorce} (Feb 2017).
\newline\urlprefix\url{shorturl.at/CEJM9}

\bibitem{wang2017fingerprint}
H.~Wang, C.~Gao, Y.~Li, Z.-L. Zhang, D.~Jin, From fingerprint to footprint:
  Revealing physical world privacy leakage by cyberspace cookie logs, in:
  Proceedings of the 2017 ACM on Conference on Information and Knowledge
  Management, 2017, pp. 1209--1218.

\bibitem{saxena2015state}
N.~Saxena, B.~J. Choi, State of the art authentication, access control, and
  secure integration in smart grid, Energies 8~(10) (2015) 11883--11915.

\bibitem{wu2016bi2g}
J.~Wu, S.~Guo, J.~Li, D.~Zeng, Big data meet green challenges: Big data toward
  green applications, IEEE Systems Journal 10~(3) (2016) 888--900.

\bibitem{wu2016big2}
J.~Wu, S.~Guo, J.~Li, D.~Zeng, Big data meet green challenges: Greening big
  data, IEEE Systems Journal 10~(3) (2016) 873--887.

\bibitem{uribe2016state}
N.~Uribe-P{\'e}rez, L.~Hern{\'a}ndez, D.~De~la Vega, I.~Angulo, State of the
  art and trends review of smart metering in electricity grids, Applied
  Sciences 6~(3) (2016) 68.

\bibitem{kumar2019smart}
P.~Kumar, Y.~Lin, G.~Bai, A.~Paverd, J.~S. Dong, A.~Martin, Smart grid metering
  networks: A survey on security, privacy and open research issues, IEEE
  Communications Surveys \& Tutorials 21~(3) (2019) 2886--2927.

\bibitem{cook2009ambient}
D.~J. Cook, J.~C. Augusto, V.~R. Jakkula, Ambient intelligence: Technologies,
  applications, and opportunities, Pervasive and Mobile Computing 5~(4) (2009)
  277--298.

\bibitem{judd_2020}
B.~Judd,
  \href{https://www.abc.net.au/news/2020-02-11/gps-tracking-watch-security-bug-data-breach-personal-info/11909478}{Smartwatch
  apps let parents keep track of their kids but data breaches mean strangers
  can watch them too} (Feb 2020).
\newline\urlprefix\url{https://www.abc.net.au/news/2020-02-11/gps-tracking-watch-security-bug-data-breach-personal-info/11909478}

\end{thebibliography}




% \end{thebibliography}

\end{document}